\documentclass[nocopyrightspace]{sigplanconf}

\usepackage{hyperref}
\usepackage{amsmath,amsthm,amsfonts,stmaryrd}
\usepackage{bbm}
\usepackage{color,xcolor}
\usepackage{listings}
\usepackage{lstcoq}
\usepackage{mathpartir}
\usepackage{graphics}
\usepackage{bussproofs}

\usepackage{tikz}

\xdefinecolor{darkgreen}{rgb}{0, 0.4, 0}

\begin{document}

\newcommand{\thetitle}{MirrorShard: Proof by Computational Reflection with Verified Hints}
\preprintfooter{\thetitle}
\title{\thetitle}


\authorinfo{Gregory Malecha}
           {Harvard University SEAS}
           {gmalecha@cs.harvard.edu}
\authorinfo{Adam Chlipala}
           {MIT CSAIL}
           {adamc@csail.mit.edu}
\authorinfo{Thomas Braibant}
           {University of Grenoble and MIT CSAIL}
           {thomas.braibant@inria.fr}
\authorinfo{Patrick Hulin}
           {MIT}
           {phulin@mit.edu}
\authorinfo{Edward Z. Yang}
           {Stanford University}
           {ezyang@cs.stanford.edu}

\maketitle

\begin{abstract}
We describe a method for building composable and extensible verification procedures within the Coq proof assistant.  Unlike traditional methods that rely on run-time generation and checking of proofs, we use verified-correct procedures with Coq soundness proofs. Though they are internalized in Coq's logic, our provers support sound extension by users with hints over new domains, enabling automated reasoning about user-defined abstract predicates.  We maintain soundness by developing an architecture for modular packaging, construction, and composition of hint databases, which had previously only been implemented in Coq at the level of its dynamically typed, proof-generating tactic language.  Our provers also include rich handling of unification variables, enabling integration with other tactic-based deduction steps within Coq. We have implemented our techniques in MirrorShard, an open-source framework for reflective verification. We demonstrate its applicability by instantiating it to separation logic in order to reason about imperative program verification. 

\end{abstract}

\section{Introduction}
\label{sec:intro}

In using proof assistants to establish theorems with very high assurance at relatively low human cost, two main methods are employed.  One can implement \emph{proof-generating theorem provers} that justify their decisions in terms of more primitive reasoning steps, or one can employ the \emph{proof by reflection} style, which involves verifying the correctness of provers formally, so that there is no need to audit their outputs.  The latter style has been used fairly heavily in the type theory community, where it is often viewed as a straightforward implementation technique to improve both performance of and assurance about proof search.  
To our knowledge, however, proof by reflection has only been applied to fairly focused problems where the domain of discourse (or its axiomatization) is known a priori. To explore its limits, we explore
mostly automated correctness verification of imperative programs, in a framework supporting higher-order logic and \emph{abstract predicates}.  \emph{Extensibility} is crucial in this domain, as the (expert) author of an abstract data type (ADT) would like to teach the prover how to reason about related proof goals that arise in verifying (non-expert) client code that calls the ADT's methods.  
We hope to achieve this \emph{modularity} and \emph{expressivity} while minimizing the overhead of proofs.

\medskip

Realistic program verification involves consideration of many details of program behavior.  Automated theorem provers play a crucial role in discharging the many straightforward proof obligations, freeing the programmer to focus on providing invariants and other pieces of high-level insight.  However, conventional automated program verifiers as in the ESC family~\cite{EscPLDI02} suffer from a number of disadvantages.
\begin{itemize}
  \item \textbf{Debuggability.} Theorem prover implementations can be quite complex.  While the prover will facilitate effective static verification for many other programs, the prover itself is usually debugged tediously via testing, and ``bugs'' often manifest as confusing failures of the tool to discharge specific goals.

  \item \textbf{Trustworthiness.} As a corollary of the last concern, we might worry that an off-the-shelf theorem prover is unsound in a way that can have serious consequences for security and so forth, if we use it as a trusted component of a verification system.
  \item \textbf{Flexibility.} The logics of automated theorem provers are generally chosen to be decidable or otherwise tractable in practice, ruling out expressive higher-order features.  As a result, some important correctness theorems (e.g., semantic correctness of a compiler) may not be possible to state, and others must be stated in prohibitively verbose ways.
  \item \textbf{Composability.} Many theorem provers act as standalone systems that do not easily produce results that can be combined with results arising from other tools using a unified logical language.  For instance, we may want to verify the correctness of a compiler (which is outside the reach of traditional program verifiers), verify the correctness of a program in the compiler's source language (which is well within the scope of traditional verifiers), and compose the results into a high-confidence deduction about the compiled version of that source program.
\end{itemize}

All of the above reasons have contributed to the popularity of program verification with proof assistants like Coq~\cite{Coq} and Isabelle~\cite{Isabelle}.  Such systems are based on small, trustworthy proof checkers for relatively simple sets of axioms and inference rules. They achieve \textbf{trustworthiness} by allowing the use of arbitrary proof \emph{search} techniques without expanding the trusted code base for proof \emph{checking}.  The underlying logics are higher order, providing very good \textbf{flexibility}.  It is even possible to leverage external tools, like SMT solvers, to generate proof assistant proofs, addressing the \textbf{composability} concern.

However, \textbf{debuggability} remains a serious concern. 
While automated provers support many features, they are not a panacea. For example, combinations of modal, linear, and higher-order logic are not supported by any prover that we are aware of.
And even when they are, the undecidability of powerful logics forces frequent use of new, problem-specific heuristics that must be embedded into automation or applied manually each time they are needed. 
Without the ability to justify such extensions, user customization compromises the trustworthiness of the entire system.
While, it is possible to use arbitrary implementations to generate proof traces in novel ways, such procedures can be nearly as hard to debug as conventional, non-proof-generating theorem provers.

It is also true that switching to an approach based on proof generation can bring substantial \textbf{performance} costs.  A proof language with few orthogonal features is very attractive because it allows proof checkers to be small and trustworthy, but small proof languages tend to promote large proofs.  Automated provers must be augmented to support generation of proofs, which adds code complexity and performance overhead.  Later, it is necessary to check the proofs, which adds further overhead.

\medskip

An established best-of-both-worlds technique for type-theoretic proof assistants is \textbf{proof by reflection}~\cite{ReflectionTACS97}.  Here, instead of writing a procedure that \emph{generates a proof} on each invocation, we instead \emph{verify the procedure itself}.  Now no proofs need to be generated, freeing us from the associated overhead.  However, the formal guarantees are just as strong, because we have proved the correctness of the procedure, which is generally implemented in a functional programming language within the proof assistant's logic.

Proof by reflection has previously been implemented for conventional mathematical decision problems.  For instance, Coq comes packaged with a tactic for algebraic simplification over mathematical rings.  This procedure, and others out there, operate over a small vocabulary of logical symbols whose semantics must be understood.  Often the procedures are \emph{extensible} in the sense of allowing customization of the set of symbols.  However, such procedures can be thought of as single-minded, never combining multiple, user-defined reasoning strategies in the way that successful automated program verifiers do.

To investigate how far the scope of proof by reflection may be expanded, we built reflective implementations of two key proof procedures for imperative program verification with \emph{higher-order separation logic}~\cite{SeplogLICS02} within the Bedrock~\cite{BedrockPLDI11} library for Coq.  Previously, these procedures were implemented in a proof-generating way, with Coq's domain-specific tactic language Ltac~\cite{LtacLPAR00}.

Listing~\ref{treeset} shows how our verification procedures handle user-defined abstractions by showing a verification of binary search tree ``lookup.''  The code defines a representation invariant for binary search trees (\coqe{bst}).
Afterward, a few \emph{hint lemmas} are proved about the invariant. These lemmas encapsulate the semantic knowledge that the reflective procedures will need to reason about \coqe{bst}.
Specifications for the imperative code make heavy use of logical quantifiers, which create additional challenges for our automation.
Listing~\ref{treeset} ends with a short proof of program correctness, via the \coqe{sep} tactic that calls the reflective procedures with a particular package of program-specific \emph{hints}.

\begin{figure}[t]
\tikzstyle{commentlabel}=[anchor=west,rotate=90,font=\small,text=darkgreen]
\tikzstyle{commentline}=[very thick,draw=darkgreen]
\begin{tikzpicture}[overlay]
  \draw[commentline] (8.05,-3.75) -- (8.05,-4.5) ;
  \node[commentlabel] at (8.25,-4.6) {user predicate};

  \draw[commentline] (8.05,-4.9) -- (8.05,-6.6) ;
  \node[commentlabel] at (8.25,-6.8) {refinement hints};

  \draw[commentline] (8.05,-7.35) -- (8.05,-8.3) ;
  \node[commentlabel] at (8.25,-8.7) {combine hints};

  \draw[commentline] (8.05,-9.9) -- (8.05,-11.1) ;
  \node[commentlabel] at (8.25,-11.6) {quantified invariants};

  \draw[commentline] (8.05,-16.9) -- (8.05,-17.5) ;
  \node[commentlabel] at (8.25,-17.6) {prove with hints};

  \draw[commentline] (3.8,-17.5) -- (4.5,-17.5) ;

\end{tikzpicture}
\vspace{-0.2cm}
\begin{lstlisting}[language=coq,caption={Verification of binary search trees implementing finite-set ``lookup''},label=treeset,captionpos=t]
(* "Spine" type to define the rep. predicate *)
Inductive tree := Leaf : tree | Node : tree -> tree -> tree.

(* Recursive representation predicate for BSTs *)
Fixpoint bst' (s : set) (t : tree) (p : W) : HProp :=
  (* details omitted *).

(* Main rep. predicate, which wraps the above with a
 * mutable pointer to its root *)
Definition bst (s : set) (p : W) := [| freeable p 2 |]
  * Ex t, Ex r, Ex junk, p =*> r * (p ^+ \$4) =*> junk * bst' s t r.

(* A standard tree refinement hint *)
Theorem nil_fwd : forall s t (p : W), p = 0 -> 
  bst' s t p ===> [| s %= empty /\ t = Leaf |].
Proof.  destruct t; sepLemma.  Qed.

(* ...more hints... *)

(* Combine the hints into a first-class package *)
Definition hints : HintPackage.
prepare (nil_fwd, bst_fwd, cons_fwd) (nil_bwd, bst_bwd, ...).
Defined.

Definition bstM : bmodule := { ...
(* Method implementation *)
bfunction "lookup"("s", "k", "tmp") [lookupS]
  "s" <-* "s";;
  [Al s, Al t,
    PRE[V] bst' s t (V "s") * mallocHeap
    POST[R] [| (V "k" %in s) \is R |] * bst' s t (V "s") * mallocHeap]
  While ("s" <> 0) {
    "tmp" <-* "s" + 4;;
    If ("k" = "tmp") {
      (* Key matches! *)
      Return 1
    } else {
      If ("k" < "tmp") {
        (* Searching for a lower key *)
        "s" <-* "s"
      } else {
        (* Searching for a higher key *)
        "s" <-* "s" + 8
      }
    }
  };;
  Return 0 ... }.

(* Prove our implementation partially correct. *)
Theorem bstMOk : moduleOk bstM.
Proof.  vcgen; abstract (sep hints; auto).  Qed.
\end{lstlisting}
\end{figure}

Our high-level contributions come from \textbf{adapting the proof-by-reflection approach to a more open-ended setting}. In particular:
\begin{itemize}
\item To the best of our knowledge, ours are the \textbf{first reflective tactics to support a notion of reusable hints}, similar to the notions exposed in Coq Ltac programming.  
Our approach leverages three types of hints to teach our core procedures about new abstract predicates.
Even the usual ``points-to'' predicate of separation logic is not built into our tactics, but rather taught to it via hints.  We identify three hint mechanisms that suffice to support all of the above from a small core, and we prove the soundness of the hint architecture. We also demonstrate (proved correct) classic data structure examples like arrays, linked lists, and search trees. 
\item We extend the proof-by-reflection approach to handle proof goals containing \textbf{quantifiers}, and we implement and verify a \textbf{unification} algorithm to facilitate related reasoning.  Coq itself includes a disjoint treatment of unification, and we have developed an approach to two-way communication between the two unification systems.  This interface is not trivial because unification is not part of Coq's logic, but rather added on outside the trusted base and interfaced with using Ltac, Coq's tactic language for building proofs.
\item Hints are naturally expressed over different logical theories, involving different data structure representation predicates and different background theories (e.g., bitvectors, lists, strings) for stating side conditions of lemmas.  We have developed a \textbf{modular architecture for composing verified hints over different theories}, including a mechanism for carving out and combining smaller domains.
\item We also provide a \textbf{performance analysis of design choices in implementation of extensible reflective tactics}. We encountered surprising challenges in achieving reasonable performance while supporting all of the above features.

\end{itemize}

We begin with background on reflective proofs (Section~\ref{sec:reflection}). We then discuss the broadly applicable novel technical devices behind our implementation, which enable us to apply reflective reasoning to the complicated expressions that arise in our verification tasks (Section~\ref{sec:extensible}).
We then try to distill the reusable engineering lessons we have learned about implementing and optimizing reflective procedures at this scale (Section~\ref{sec:lessons}).
Next comes an evaluation of the automation achieved by our procedures and an analysis of their performance characteristics compared to non-reflective verification (Section~\ref{sec:evaluation}), where our overall conclusion is that we improve asymptotic performance substantially, though well-abstracted programs and specifications may not lead to large enough invariants to exhibit the scaling improvements clearly.  We wrap up with a discussion of related work (Section~\ref{sec:related-work}).

Our techniques, and many of the components that we built, are implemented in the MirrorShard library, available for download at:
\begin{center}
  \url{https://github.com/gmalecha/mirror-shard/}
\end{center}

\section{A Primer on Proof By Computational Reflection}
\label{sec:reflection}

In this section, we discuss the idea of reflection~\cite{ReflectionTACS97}. 
For the sake of simplicity, we shall present 
reflection using a stripped-down example: here, we are interested in
reducing the proof of equalities like $f~a~(g~b) = f~c~(g~(h~a))$ to
the proof of $a = c$ and $b = h~a$. That is: we are computationally
discharging equalities between terms in the (multi-sorted) algebra
generated by an arbitrary signature, generating new proof obligations for
equalities between subterms that do not follow from basic properties of
equality. 

The first step of reflection is to
\emph{encode the syntax of proof goals in a datatype defined within your
proof assistant's logic}.  In the case of Coq, this logic is \emph{Gallina},
a dependently typed lambda calculus with inductive definitions.
Thus, we start by defining a datatype \coqe{expr} to represent terms: in
this syntaxified representation, an expression is just a function symbol
applied to a list of expressions (see Listing~\ref{lst:reif}). We break the
circularity using 0-argument functions to represent variables and constants.
This is important because it is imperative that the \coqe{expr} type has a
decidable equality so we can avoid generating obligations such as \coqe{x = x}.
We achieve this by representing functions as indices into an environment of
functions.
%

To make the meaning of our syntax formal, we define a denotation function
\coqe{denote} that maps an expression \coqe{e} supposedly of the type
represented by \coqe{ty}, to a value of type \coqe{option (nth types ty)}.
Note that this denotation function is partial, because our data type admits
the encoding of ill-typed terms. (We will return to this particular design
choice in Section~\ref{sec:lessons}.) 
The denotation function is parameterized by an environment of types and
function signatures (of type \coqe{sig} indexed by the type environment),
and will perform dynamic type-checking to ensure that the Gallina term
that it produces is well-typed.
Thus, where \coqe{ltb} is a Boolean-valued less-than test for natural numbers, the term \coqe{ltb (x + y) z} can be represented using the following environments and term.
\vspace{0.2cm}
\begin{coq}
Let types := [nat; bool]. 
Let functions := [sig [0;0] 0 plus; sig [0;0] 1 ltb; 
                  sig [] 0 x; sig [] 0 y; sig [] 0 z]  

Func 1 [Func 0 [Func 2 []; Func 3 []]; Func 4 []] : expr
\end{coq}

Using this representation, we can \emph{implement a (heuristic) decision procedure} in Gallina. The procedure considers the head symbols of two
expressions. If they are the same, it proceeds
recursively on their arguments. Otherwise it accumulates a ``proof obligation''
(i.e., a pair of expressions whose denotations must be proven equal).
\begin{coq}
Fixpoint f_eq (a b: expr) (ty: nat) : list (nat*expr*expr) :=
  match a , b with
    | Func f1 args1 , Func f2 args2 =>
      if f1 == f2
      then union (map3 f_eq args1 args2 (domainOf f1))
      else [(ty, a, b)]
  end.
\end{coq}

Finally, we prove the procedure sound. That is, we prove that if all of the constraints are satisfied
then the denotations of the original terms are equal.
\begin{coq}
Theorem f_eq_correct : forall a b ty,
  Forall (fun (t,x,y) => denote x t = denote y t) (f_eq a b ty) 
  -> denote a ty = denote b ty.
\end{coq}

\begin{figure}
  \begin{lstlisting}[frame=none,caption={Representing multi-sorted expressions},label=lst:reif]
Inductive expr :=  
| Func: nat -> list expr -> expr.

Variable types : list Type.

Fixpoint ftype (domain: list nat) (range: nat) : Type :=
  match dom with 
    | [] => nth range types
    | t::dom => nth t types -> ftype dom range
  end. 

Record sig := {dom: list nat; rng: nat; val: ftype dom rng}. 

Variable functions : list sig.
Fixpoint denote (e: expr) (ty: nat): option (nth ty types) := ...
\end{lstlisting}
\end{figure}

Applying this lemma makes it possible to replace several proof steps
(one proof step for every common head symbol between the lefthand-side
and righthand-side expressions) by a single proof step, plus a
\emph{computation}.  That is, we come to the final crucial element of
a proof by reflection: \emph{a goal is proved by appealing to a theorem
and then proving its hypothesis by ``running'' the hypothesis
to reduce it to a normal form, which should then be much easier to prove
than the original goal.}

For example, suppose that we want to prove %
\coqe{ltb (x+y) z = ltb (x+y) w}. We can apply the aforementioned
lemma, using suitable values \coqe{a} and \coqe{b} of type
\coqe{expr}, and Coq will check that \coqe{denote 1 a} (resp.
\coqe{denote 1 b}) is convertible to the lefthand side (resp.
righthand side) of the goal, according to the lambda calculus reduction
rules of Gallina.  Here we take advantage of the fact that, in Coq's logic,
reduction-equivalent terms may always be used interchangeably, with no
need to include explicit proof steps as justification.

\paragraph{Reification} In the above discussion, we side-stepped a
difficulty: we have to automate the construction of the syntactic
representation (i.e., terms of type \coqe{expr}) from the terms that
appear in the goal. While this operation, called \emph{reification}, is conceptually the dual of
\coqe{denote}, it must be performed at the meta level using special-purpose tactics.
We shall return to this problem in Section~\ref{sec:lessons}.

\section{An Extensible Verification Architecture}
\label{sec:extensible}

Compared to past work on proof by reflection, our verification architecture achieves its powerful form of extensibility through three novel technical devices:
\begin{description}
\item[Extensible Syntax] enables us to encode terms with arbitrary Coq constants and quantifiers, while retaining the ability to compute on constants embedded in terms. (Section~\ref{sec:representation})
\item[Composable Soundness] enables us to combine soundness proofs about procedures that reason about different logical domains. (Section~\ref{sec:composition})
\item[Integration with Unification] enables flexible integration of reflective procedures with traditional Ltac-based automation that uses unification variables that are not formalized in Coq's core logic. (Section~\ref{sec:unification})
\end{description}

\subsection{An Extensible Syntax with Binders}
\label{sec:representation}
Our reified syntax for assertions of separation logic~\cite{SeplogLICS02} (summarized in Listing~\ref{lst:syntax}) is similar in spirit to the generic syntax described in Section~\ref{sec:reflection} but incorporates several new forms for dealing with binders, manipulating proofs, and performing domain-specific reasoning. The four key differences are:
\begin{figure}
\begin{lstlisting}[frame=none,caption={Representing separation logic expressions with binders},label=lst:syntax]
Record type := { Impl : Type ; Eq : Impl -> Impl -> bool }.

Inductive tvar := tvProp : tvar | tvType : nat -> tvar.

Inductive expr (ts : list type) : Type :=  
| Const:  forall (ty : tvar), tvarD ty ts -> expr ts
| Var: nat -> expr ts
| UVar : nat -> expr ts
| Func: nat -> list (expr ts) -> expr ts
| Equal : expr ts -> expr ts -> expr ts.

Inductive sexpr (ts : list type) : Type :=
| Star : sexpr ts -> sexpr ts -> sexpr ts
| Emp : sexpr ts
| Pred : nat -> list (expr ts) -> sexpr ts
| Inj : expr ts -> sexpr ts
| Exists : tvar -> sexpr ts -> sexpr ts.

Fixpoint exprD (ts : list type) (fs : list (signature ts))
  (vars uvars : list { t : tvar & tvarD ts t })
  (e : expr ts) (t : tvar) : option (tvarD ts t) := ....

Fixpoint sexprD (ts : list type) (fs : list (signature ts))
  (ps : list (psignature ts))
  (vars uvars : list { t : tvar & tvarD ts t })
  (e : sexpr ts) : hprop := ....
\end{lstlisting}
\vspace{-0.3cm}
\end{figure}
\begin{description}
\item[A distinguished encoding for the type of logical propositions] (\coqe{tvProp}) simplifies representing logical properties and enables us to represent polymorphic equality (\coqe{Equal}), even though our encoding does not support polymorphic functions in general.
\item[An expression type family] indexed by a type environment, which gives for each type both its \emph{Coq representation} and a compatible \emph{equality testing function}.  The \coqe{Const} constructor is used to inject terms that our unification algorithm should consider comparing for equality by calling type-specific procedures, in contrast to the syntactic equality check used for the rest of the constructors.
\item[Binders and local variables] are represented by the constructors \coqe{Exists} of \coqe{sexpr} and \coqe{Var} of \coqe{expr}.  Our encoding is similar to the \emph{locally nameless} technique for lambda terms, in that we maintain distinct representations of global/free variables (nullary function applications via \coqe{Func}) and local/bound variables (via \coqe{Var}).
\item[Unification variables] (represented with \coqe{UVar}) are represented explicitly so that our procedures may deduce and substitute values for them.  Informally, a \coqe{Var} expression supports universal-quantifier reasoning, while a \coqe{UVar} expression supports existential-quantifier reasoning.  We must prove any theorem considering all possible \coqe{Var} values, but we are allowed to choose specific \coqe{UVar} values that make the theorem true.
\end{description}
To support our new features, the denotation functions \coqe{exprD} and \coqe{sexprD} have several new parameters for variables, unification variables, and separation logic predicates.

To see how these components fit together, we show a simple heap assertion for a cons cell where the first value is a unification variable from the context ($\mathsf{?a}$) and the second (occurring 4 bytes later) is an existentially quantified value ($v$):
\vspace{-0.05cm}
$$
\exists v, \mathsf{p} \mapsto \mathsf{?a} * (\mathsf{p} + 4) \mapsto v
$$
\vspace{-0.06cm}
This term can be represented as:
\vspace{-0.06cm}
\begin{coq}
Let types := [(word, eq_word)].
Let funcs := [([],tvType 0,p); ([tvType 0;tvType 0],tvType 0,+)].
Let preds := [([tvType 0; tvType 0], $\mapsto$)].
Let vars  := []. (* p is represented as a function *)
Let uvars := [(tvType 0, ?a)].
Exists (tvType 0)
        (Star (Pred 0 [Func 0 [] ; UVar 0]) 
              (Pred 0 [Func 1 [Func 0 []; Const 4]; Var 0]))
\end{coq}

\subsection{Achieving Compositionality}
\label{sec:composition}
To be useful, hints must be both \textbf{self-contained}, packaging together the hint and its soundness proof, and \textbf{compositional}, enabling us to combine separately defined hints in a meaningful way. Here, our type-family representation introduces problems. The most na\"ive prover type, with one environment of types fixed for all provers, will never compose with provers using different types. We solve the problem using a sort of universal quantification over type environments that imposes constraints on the presence of specific identifiers.

\paragraph{A Constraint Formulation}
One way of representing the constraints is propositionally, with explicit logical assertions. We can encode constraints using partial environments and say that an environment ($e$) satisfies a constraint ($C$), written $C \models e$, when all mappings in the constraint are consistent with the environment. Using this formulation, we can define two provers, one for lists and the other for machine words.  Each constraint is a list of optional types, where the presence of a type forces the final environment to contain that type in that list position.
\begin{coq}
Let C1 := [None; Some (list word)].
Definition prover_1 : forall ts, C1 $\models$ ts -> expr ts -> bool.
Let C2 := [Some word].
Definition prover_2 : forall ts, C2 $\models$ ts -> expr ts -> bool.
\end{coq}

Since \coqe{C1} and \coqe{C2} are compatible, these provers can be composed into a new prover that accepts an environment that satisfies both \coqe{C1} and \coqe{C2}. The difficulty of this formulation arises in type-checking prover implementations, where it is generally necessary to use \emph{casts} justified by appealing to the consistency proof.  For example, suppose that \coqe{prover_2} is determining whether a number is a multiple of 4. An ideal formulation of the soundness of this prover would be:
\begin{coq}
Theorem prover2_sound' : forall ts fs vars uvars e v (pf : C2 $\models$ ts),
  FC2 $\models$ fs -> prover_2 ts pf e = true ->
  exprD ts fs vars uvars (tvType 0) e = Some v ->
  v mod 4 = 0.
\end{coq}
where \coqe{FC2} is \coqe{prover_2}'s constraint on the function environment. Unfortunately, this soundness statement is not well typed. The problem stems from \coqe{v}. In line 3, \coqe{v} must have type \coqe{tvarD ts (tvType 0)}, while in line 4 \coqe{v} must have type \coqe{word}.

The core problem lies in the intensional nature of Coq's type theory. Coq distinguishes between two notions of equality.
\begin{description}
\item[Definitionally equal ($\equiv$)] terms are identical after reduction. This notion of equality is part of Coq's core logic and requires no extra work to apply during type-checking.
\item[Provably equal ($=$)] terms are defined by a binary inductive predicate \coqe{x = y}. This type encodes an explicit proof (in Coq's logic) that \coqe{x} and \coqe{y} are equal. To use this type of equality, we must perform a cast using the proof.
\end{description}
For a concrete example, we return to the problem above. From the meaning of $\models$ we can prove \coqe{tvarD ts (tvType 0) = word}, but the two are not definitionally equal under Coq's reduction rules, since \coqe{ts} is a variable and not a concrete environment; the reduction to determine equality gets stuck examining the structure of \coqe{ts}. We can solve this problem using the following cast:
\begin{coq}
Theorem prover2_sound : forall ts fs vars uvars e v (pf : C2 $\models$ ts),
  FC2 $\models$ fs -> prover_2 ts pf e = true ->
  exprD ts fs vars uvars (tvType 0) e = Some v ->
  (cast v (GetConsistent pf 0)) mod 4 = 0.
\end{coq}
where \coqe{GetConsistent} takes the consistency proof and the index and returns a proof of \coqe{tvarD ts (tvType 0) = word} (i.e. by looking up the index in the constraint).  The cast is present for a reason: removing it produces an ill-typed term.  Therefore, it would be unsound to include a reduction rule that removes useful casts.  The problem is that proving triviality of casts (i.e., that they convert between \emph{definitionally} equal terms) is extra work with no counterpart in pencil-and-paper reasoning.  A proof step like this one must be preceded by another proof step that rearranges the proof context to a form that is \emph{well-typed both before and after the cast is removed}, which can be surprisingly subtle and case-specific.  Furthermore, justifying this step (``removing a cast between definitionally equal terms has no effect'') is not possible in Coq's core logic without appealing to axioms. 

\vspace{0.45cm}
\paragraph{The Computational Formulation}
\label{sec:computational}
While Coq's reduction mechanism does not handle the above formulation well, we can give an alternative formulation that enjoys better computational properties. In this formulation we achieve constrained quantification over environments not by starting from an arbitrary environment and asserting a constraint over it, but rather by \emph{starting with an arbitrary environment and performing a computation on it to make it constraint-compliant by construction}.  The heart of the technique is a recursive function called like \coqe{applyC c e}, which ``instruments'' environment \coqe{e} to satisfy constraint \coqe{c}.
\begin{coq}
Fixpoint applyC (c: constraint T) (e: list T) : list T :=
  match c with 
    | nil => e
    | None :: c' => hd d e :: applyC c' (tl e)
    | Some v :: c' => v :: applyC c' (tl e)
  end.
\end{coq}
Reformulating the previous theorem leads to the following: 
\begin{coq}
Definition prover_2 : forall ts, 
  let ts' := applyC C2 ts in expr ts' -> bool.

Theorem prover2_sound : forall ts fs vars uvars e v,
  let ts' := applyC C2 ts in let fs' := applyC FC2 fs in
  exprD ts' fs' vars uvars (tvType 0) e = Some v ->
  prover_2 ts' e = true ->
  v mod 4 = 0.
\end{coq}

Now, \coqe{tvarD (applyC C2 ts) (tvType 0)} is definitionally equal to \coqe{word} since the environment has been reformed by \coqe{applyC} so that it is manifestly a series of ``cons'' operations, containing the proper type in the proper position.  In particular, reduction tells us \coqe{applyC C2 ts $\equiv$ word :: tl ts}.  In general, we may now extract any constant index occurring in the constraint from the updated environment, without any need for dependent casts. 

In fact, this formulation gives us much more: it enables \emph{computational} composition. When two constraint environments are compatible, i.e. they do not specify different values for any index, applying \coqe{applyC} to them commutes \emph{definitionally}.
$$
\mathsf{applyC}\, C1\, (\mathsf{applyC}\, C2\, e) \equiv 
\mathsf{applyC}\, C2\, (\mathsf{applyC}\, C1\, e)
$$
This feature makes composition of two provers trivial. If we wish to compose two provers, say \coqe{p1} and \coqe{p2}, with different, but compatible, environments \coqe{TC1} and \coqe{TC2}, then we can pre-compose each function by applying the other's type constraint to produce two provers on the same environment. Concretely:
\begin{coq}
(fun ts => p1 (applyC TC2 ts)) 
   : forall ts, expr (applyC TC1 (applyC TC2 ts)) -> bool
(fun ts => p2 (applyC TC1 ts)) 
   : forall ts, expr (applyC TC2 (applyC TC1 ts)) -> bool
\end{coq}
Since these types are definitionally equal, we can treat them identically, applying both functions to the exact same term.  That is, any composition operation can be written free of both \emph{explicit proofs} and \emph{explicit casts}.

Using this representation, we can package together reusable, self-contained verification hints using Coq's dependent records.
\begin{coq}
Record HintDatabase :=
{ Types  : constraint type
; Funcs  : forall ts, constraint (signature (applyC Types ts))
; Preds  : forall ts, constraint (psignature (applyC Types ts))
; Hints  : forall ts, HintsT (applyC Types ts)
; Hints_correct : forall ts fs ps, 
  HintsT_correct (Hints ts) (applyC (Funcs ts) fs) 
                             (applyC (Preds ts) ps) }.
\end{coq}
The first three fields express the constraints on the type, function, and heap predicate environments. The fourth field contains a record that packages together the three different types of hints that our system uses (discussed in Section~\ref{sec:procedures}). Note that the \emph{type} of \coqe{Hints} is dependent on the \emph{value} of the \coqe{Types} field. The final field encapsulates the soundness proofs for the hints. When applying our reflective procedures, our tactics use the first three fields to seed the environments used to reify terms, the hints to compute the results, and the soundness proof to justify reasoning with the hints. 

Because we use a shallow embedding of constraints, we do not need to write a proof that two packages compose. Unfortunately, this means that we cannot write a Gallina function that combines two hint databases. Rather, we use Ltac (a dynamically typed language) to construct the term and turn the type checker loose on it. If the result type-checks, then the environments are consistent and the hints compose; otherwise, the programmer gets an error message about compatibility.

\subsection{Interfacing with Unification Variables}
\label{sec:unification}

Traditional proofs in Coq are done in ``proof mode'' using tactics that manipulate a goal that is displayed to resemble a standard sequent calculus. Universally quantified variables are displayed in a ``proof context'' above a double line, while the goal is displayed below the line. In order to integrate well with existing tactic-based and interactive proof techniques, our reflective procedures must fit naturally into this view.

\begin{figure}
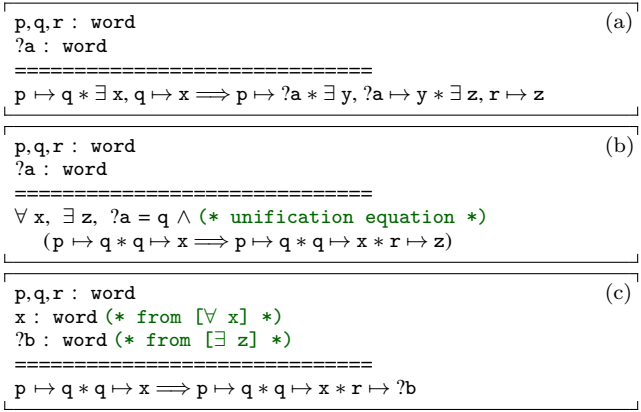

     \begin{lstlisting}[gobble=6,language=coq]
       p,q,r : word                                                  $\mathrm{(a)}$ 
       ?a : word
       ==============================
       p $\mapsto$ q * exists x, q $\mapsto$ x ===> p $\mapsto$ ?a * exists y, ?a $\mapsto$ y * exists z, r $\mapsto$ z
     \end{lstlisting}
     \vspace{-0.2cm}
     \begin{lstlisting}[gobble=6,language=coq]
       p,q,r : word                                                  $\mathrm{(b)}$ 
       ?a : word 
       ==============================
       forall x, exists z, ?a = q /\ (* unification equation *)
          (p $\mapsto$ q * q $\mapsto$ x ===> p $\mapsto$ q * q $\mapsto$ x * r $\mapsto$ z)
     \end{lstlisting}
     \vspace{-0.2cm}
     \begin{lstlisting}[gobble=6,language=coq]
       p,q,r : word                                                  $\mathrm{(c)}$
       x : word (* from [forall x] *)
       ?b : word (* from [exists z] *)
       ==============================
       p $\mapsto$ q * q $\mapsto$ x ===> p $\mapsto$ q * q $\mapsto$ x * r $\mapsto$ ?b
     \end{lstlisting}

  \caption{Representation of variables as they pass through our verification procedures: (a) initial goal; (b) direct output of the unification procedure; (c) after simplification with Ltac}
  \label{fig:unification-variables}
\end{figure}

Figure~\ref{fig:unification-variables} demonstrates how our reflective procedures manipulate binders and unification variables. The implementation of these procedures is complicated by manipulation of de'Bruijn indicies but is otherwise mostly standard. It is the phrasing of the soundness theorems that we focus on here. We begin with an illustrative example, Figure \ref{fig:unification-variables}.(a), which shows a simple heap implication (denoted by $\Longrightarrow$) with three internal quantifiers and a unification variable (\coqe{?a}). Unlike in normal Coq output, we include unification variables explicitly in proof contexts, i.e. we can pick any term of type \coqe{word} for \coqe{?a} as long as it mentions only globals and variables that occur above it.

Our procedures take as inputs goals, like the one in Figure \ref{fig:unification-variables}.(a), after they are reified as terms in the logic.  For explanatory purposes, here we consider a simple procedure that only performs unification, attempting to learn the values of both bona fide Coq unification variables (e.g., \coqe{?a}) and variables that are quantified existentially in the conclusion of the implication (e.g., \coqe{y} and \coqe{z}).  A key question is how the unification procedure, a pure function in the logic, can \emph{cause side effects} to resolve unification variables in the original Coq proof context.  As in proof by reflection in general, our procedures may only announce results by replacing one logical formula with another that has been proven to imply the original.

The result of the unification procedure is shown in Figure \ref{fig:unification-variables}.(b).  Four different sorts of variables have been handled in four different ways.  First, a variable \emph{existentially quantified on the lefthand side of the original implication} (e.g., \coqe{x} here) is returned via a top-level universal quantification.  Second, a \emph{normal Coq unification variable} (e.g., \coqe{?a} here) is asserted to be equal to whatever replacement has been inferred for it.  Finally, there are two cases for a variable \emph{existentially quantified on the righthand side of the implication}.  Either no unification was found for it (e.g., \coqe{z} in this example), in which case it gets a top-level existential quantifier in the new goal; or some unification was found (e.g., \coqe{y} in this example) and the variable is simply removed by substituting for it everywhere it appears.

A unification is represented as a map from unification variable to syntactic expression. As for our other syntactic representations, our soundness proofs ascribe a denotation to unifications, this time as a conjunction of provable equalities (\coqe{substD}). Using this denotation function in a premise, we can prove that \emph{syntactic} instantiation, by \coqe{instantiate}, preserves the \emph{semantic} meaning of terms.
\begin{coq}
Theorem substD_instantiate : forall funcs U G e t sub,
  substD funcs U G sub ->
  exprD funcs U G (instantiate sub e) t = exprD funcs U G e t.
\end{coq}

The final step occurs in Ltac. We simplify the goal by moving variables ``above the line'' into the proof context.
All \coqe{forall} quantifiers lead to normal Coq variables (e.g., \coqe{x}), and all \coqe{exists} quantifiers lead to Coq unification variables (e.g., \coqe{?a}).  More importantly, we remove the unification equations like \coqe{?a = q} by first performing the \emph{side effect} of setting \coqe{?a} equal to \coqe{q} and then proving the equation trivially by reflexivity.  These side effects are possible in Ltac but not Gallina, and our strategy for generating initial output goals is designed to be very telegraphic in suggesting side effects. 

\section{Reflective Procedures}
\label{sec:procedures}

\begin{figure}
  \centering
  \scalebox{0.4}{\includegraphics{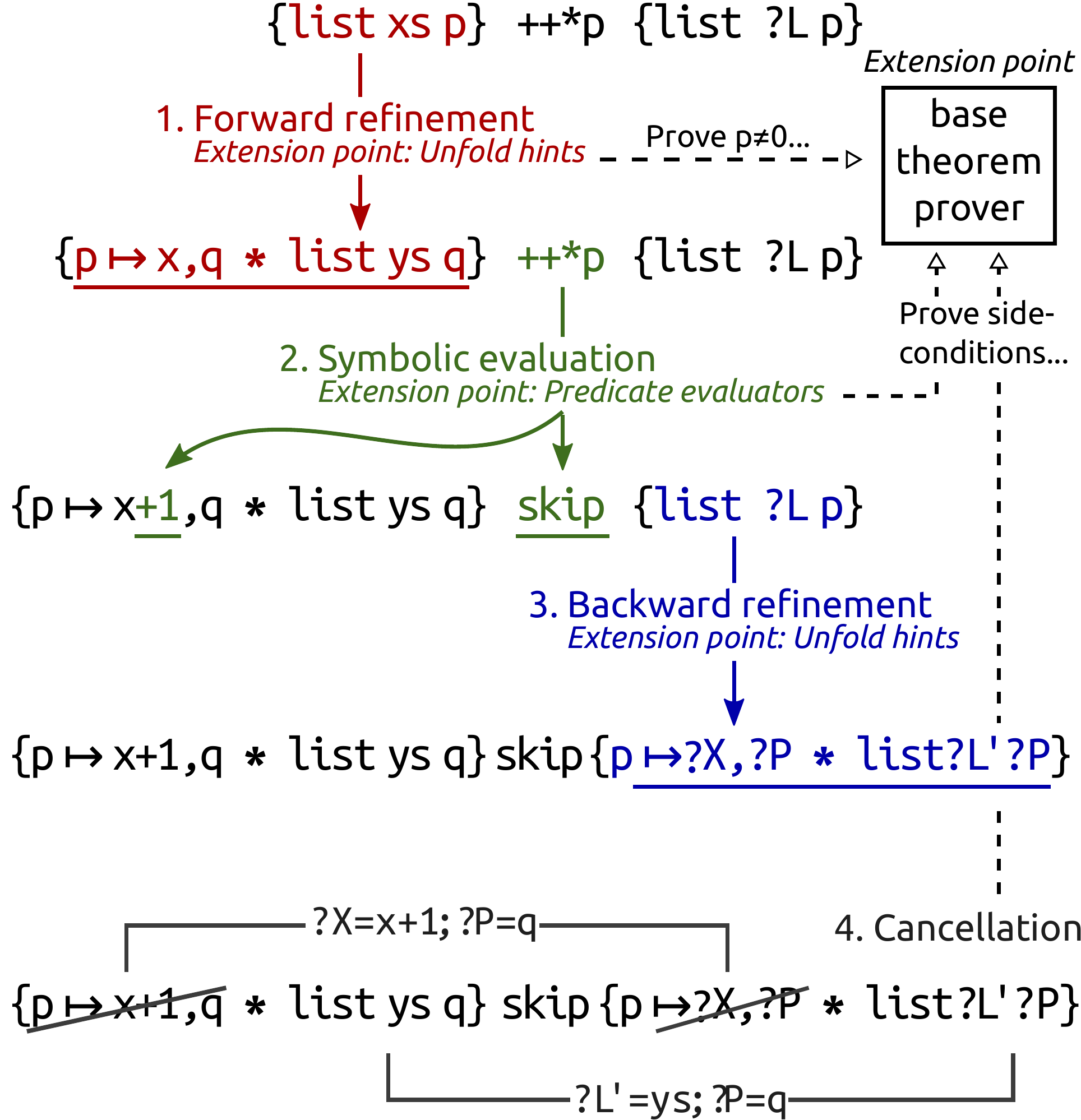}}
  \caption{The high-level verification strategy applied to a simple program that increments a memory cell}
  \label{fig:strategy}
\end{figure}

The techniques described in the previous sections enable extensible reflective verification with rich formulas. In this section, we show how to apply MirrorShard, the Coq library we have built with those techniques, to create reflective automation for the Bedrock~\cite{BedrockPLDI11} library, which supports program verification in separation logic. Our development leverages MirrorShard to build the two core verification components: symbolic execution and separation logic cancellation. The procedure is illustrated in Figure~\ref{fig:strategy} and described in the rest of the section. 

\paragraph{Symbolic Execution}
First, symbolic execution takes a precondition and a sequence of instructions and computes a post-condition. Symbolic execution starts by using pure facts in the context to refine abstract predicates. In the example, the procedure finds that the linked list predicate can be unfolded since the head pointer is not null. This information is encoded in our first type of hint:
\begin{description}
\item[Refinement Hints] are stylized Coq theorems that express predicated heap implications. These theorems are reified into Coq as inhabitants of a record type with fields for a list of universally quantified variables, a list of pure premises (\coqe{expr}s), and the expressions on each side of the separation logic implication. Due to their first-order nature, these hints can be constructed completely automatically (using Ltac programs) from actual Coq theorems that have a particular syntactic form. 
\end{description}
Pure facts, such as the side conditions on refinement hints (e.g., ``pointer is not null''), are discharged by our second type of hint:
\begin{description}
\item[Base Theory Provers] are verified Coq functions taking the place of Coq tactics (since we cannot call Ltac from Gallina code), proving \coqe{expr}-encoded proof goals with arbitrary algorithms that can be coded and verified in Gallina.

We have developed four provers:
\begin{itemize}
\item A \textbf{reflexivity prover}, which proves statements of the form $e = e$.
\item An \textbf{assumption prover}, which maintains a list of known facts and attempts to find the goal as a syntactic match to one of these facts.
\item A prover for reasoning about \textbf{linear arithmetic on width-32 bitvectors} to prove equalities and inequalities. This prover makes inferences by combining hypotheses representing expressions $e_1 = e_2 + k$ for constants $k$.  (This last form of reasoning is especially applicable to common patterns of pointer arithmetic.)
\item A prover oriented toward \textbf{array bounds checks}, which understands that array writes preserve length.
\end{itemize}
We support composing provers in a simple disjunctive style, i.e. a proposition is provable if either of two provers can prove it. 
\end{description}
After predicate refinement, symbolic execution begins interpreting instructions. Total arithmetic and logical instructions are trivial to model by converting the transfer functions into their syntaxified forms. Instructions that access memory (both reading and writing) require more care, but our use of separation logic enables an effective algorithm based on our third type of hint:
\begin{description}
\item[Memory Evaluators] are verified functions that reason about reads from and writes to heaps satisfying a separation logic assertion.
This approach enables the symbolic evaluator to interpret memory operations in terms of many different data structure predicates, without the need to expose individual points-to assertions algebraically.

In addition to a composition operator that combines two memory evaluators into one, we have implemented memory evaluators for 32-bit points-to, arrays (of both words and bytes), and local variable stack frames.
\end{description}
In the example, the memory evaluator uses the provers to inspect the separation logic formula and determines that the value read from \coqe{p} is \coqe{x} (because a subformula \coqe{p =*> x} appears) and the value written is therefore \coqe{x+1}, which is constructed syntactically when interpreting the addition.


\paragraph{Cancellation}
Cancellation proves that the strongest post-condition, computed by symbolic execution, implies the specification's post-condition. The algorithm begins with backward refinement, which uses the same type of refinement hints but refines in the conclusion of the heap implication~\footnote{Forward and backward refinement are both thin wrappers around a common, general, unification-based procedure for determining when quantified-equality hints apply to a goal.}. 
In the example, an analogous hint refines the \coqe{list} predicate, exposing the first cell. In the conclusion, existential variables are introduced as new unification variables that the rest of cancellation will attempt to instantiate.

The core of the cancellation algorithm uses the cancellative properties of separation logic to prove the implication. Since cancellation leverages unification to resolve unification variables and does not backtrack, the order of considering predicates to cancel matters. We use a simple heuristic based on a lexicographic ordering of syntactic expressions where unification variables have the highest values and are thus unified last. This ordering, for example, will attempt to unify \coqe{p $\mapsto$ ?a} with \coqe{p $\mapsto$ v} before it tries to unify it with \coqe{?b $\mapsto$ ?c}. 

The Bedrock-specific instantiation of MirrorShard, including all of the examples, is available online:
\begin{center}
  \url{https://github.com/gmalecha/bedrock-mirror-shard}
\end{center}

\section{Lessons Learned: Engineering Reflective Proof Procedures in Coq}
\label{sec:lessons}

In this section we highlight a variety of design choices that arise in the development of reflective decision procedures. While the details are Coq-specific, the ideas generalize and shed light on interesting design decisions both for users and developers of proof assistants.

\subsection{Term representation}  
General dependent types provide many representation alternatives that are not available in most programming languages. The first implementation choice is whether the type of terms should guarantee that every term is well-typed. Such a representation allows us to make the denotation function (\coqe{exprD}) total,  simplifying theorem statements and avoiding the need to prove that functions preserve the well-typedness of terms.

The cost of this convenience is indexing terms by additional environments. In our setting, we would need to parameterize \coqe{expr} by the environments of functions, variables, and unification variables in addition to the expression type, leading to a type like:
\begin{coq}
Inductive dexpr (ts : list type) (fs : list signature)
   (uvars vars : list tvar) : tvar -> Type := ...
\end{coq}
This representation moves the hard work from the soundness proofs to the computational operations that manipulate terms. For example, with dependent types, we weaken terms by structural recursion:
\begin{coq}
Fixpoint dexpr_weaken ts fs u u' g g' t (e : dexpr ts fs u g t) 
  : dexpr ts fs (u ++ u') (g ++ g') t := ...  
\end{coq}
Only the variable cases are interesting, essentially needing to justify that valid references into \coqe{u} (respectively \coqe{g}) are the same as references into \coqe{u ++ u'} (respectively \coqe{g ++ g'}).

Using a non-dependent representation, this function becomes a no-op since environments are extended at the end. Our proofs appeal to the following lemma that relates the meanings of expressions to their meanings under weakened environments.
\begin{coq}
Theorem exprD_weaken : forall ts fs u g t e v,
  exprD ts fs u g e t = Some v ->
  exprD ts fs (u ++ u') (g ++ g') e t = Some v.
\end{coq}

\begin{figure}
  \centering
  \scalebox{0.6}{\includegraphics{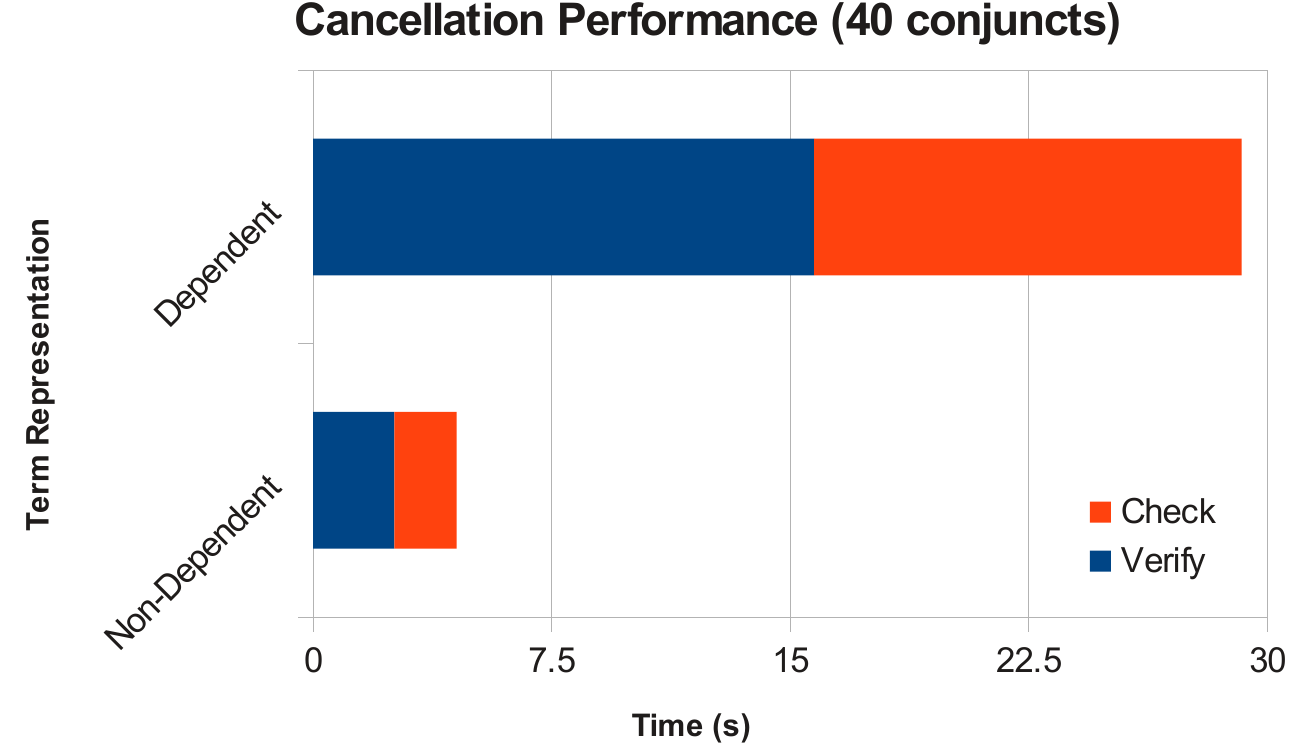}}

  \caption{Verification times for two term encodings}
  \label{fig:perf}
\end{figure}

This ease of implementation also translates to performance improvements for procedures like cancellation, as shown in Figure~\ref{fig:perf}. The first bar uses the dependent representation while the second uses the minimally dependent representation described in Section~\ref{sec:representation}. We believe that the large difference between the bars is due to the need to evaluate proof terms to reduce dependent casts, though Coq provides no profiling tools for verifying this hypothesis.

\subsection{Efficient Computation}
Analogous to the choice of term representation is the choice of function implementation. We may implement functions with dependent types, making their properties manifest directly; or we may choose simple types and then prove properties after the fact. The most common manifestation of this choice is for equality-testing functions. In the dependent style, it is common to write the type of an equality function as: \coqe!forall x y, {x = y} + {x <> y}! (which is a particular case of dependent sum type carrying proofs). The non-dependent version (\coqe{eqb}) returns a Boolean, and provides a separate proof: \coqe{forall x y, eqb x y = true $\leftrightarrow$ x = y}.

From a computational perspective, the latter is much more efficient under call-by-value reduction, since in the former the proofs that are constructed must be reduced completely. Changing our algorithms to use the non-dependent version of equality checks resulted in a 40\% reduction in the proof generation and checking time. (Note also that in the context of code extraction from Coq, the difference between the solutions disappears.)

While this fact may be well-known for seasoned implementers of reflective decision procedures, there is little guidance from Coq's standard library toward this choice. In adapting the proofs, we have found it easy to recover the proof behavior of the dependent equality using Coq's dependent type classes~\cite{sozeau08typeclasses}. 

Traditional type classes carry additional information about \textit{types}, e.g. an equality decider. Dependent types enable type classes to carry additional information about \textit{values}. For example, a type class indexed by a function can carry a proof about that function:
\begin{coq}
Class EqOk (T : Type) (f : T -> T -> bool) : Type :=
{ eq_ok : forall x y, f x y = true $\leftrightarrow$ x = y }.
\end{coq}
Proofs and automation can now reference the symbol \coqe{eq_ok} and Coq's type class resolution will attempt to find an appropriate instance. This approach is similar to the development in the math classes project~\cite{math-classes} and is the core principle underlying recent work on ExtLib~\footnote{\url{https://github.com/coq-ext-lib/coq-ext-lib}}.

\subsection{Reification}
Most sources gloss over the (mostly) uninteresting problem of reifying terms. While not particularly glamorous, the reification process can dramatically affect verification time. 

Our first version of reification for pure expressions, separation logic formulas, and object-language commands used Ltac. However, an initial performance evaluation showed that reification was a major performance bottleneck (taking almost 50\% of total verification time). 
First, some of this overhead can be attributed to Ltac itself: the language is dynamically typed and built for writing backtracking tactics for proof search, rather than building actual terms. 
Second, we had to circumvent the lack of support for manipulating open terms (i.e., terms with free variables) caused by embedded binders. The trick requires copious use of second-order pattern matching, which is considerably more time-intensive and results in code that is more difficult to read and maintain. 
Third, we had to split reification into two passes: first, to gather the type environment, and then to build the reified terms indexed by this environment. 

To address the performance problem, we implemented a second version of our reification as an OCaml plugin. This alleviates all of the previously mentioned problems: we use OCaml data structures (which are more efficient than their Ltac counterparts) to build environments and terms; we manipulate open terms, which makes reification more direct; we make a single pass on the Coq term to build the reification environment and the reified terms, rather than two in Ltac.
This plugin does not need to be trusted more than any Ltac tactic since it constructs terms that are fed into the Coq kernel. 

Using this plugin dramatically reduced the time spent on reification.  Whereas Ltac reification had taken approximately 32 minutes for a test suite of 10 examples, OCaml reification takes approximately 22 seconds (an 88X speedup). 
Processing of a single file will typically perform hundreds of reifications, so these figures are quite reasonable.  

The reason for this slowdown is more than the interpretive overhead of Ltac.  The explanation is a Coq feature (misfeature?) dealing with building terms in Ltac. 
To illustrate, consider two different Ltac expressions that build a natural number by repeatedly applying the successor function \coqe{S} to the zero constant \coqe{O}.  First, there is the simple version \coqe{S (S (...O...))}, which just builds the term directly.  For reasonably sized numbers, this expression evaluates instantly.  Then, there is the expanded form \coqe{let n := O in let n := (S n) in ...(S n)...}, which binds an Ltac-level variable for each intermediate term.  This expression evaluates \emph{in time quadratic in the term size}.

The underlying problem is that \emph{Coq re-typechecks all parts of each new term that is constructed in Ltac}.  This is not an unreasonable-sounding requirement for a dynamically typed language.  Ltac-bound variables appearing in Gallina terms have their contents substituted in explicitly.  Reification naturally builds terms step-by-step through a recursive process.  The overhead of repeated typechecking can become overwhelming in such cases.  In contrast, all Coq-level type checking in OCaml 
is done explicitly. This allows dramatic speed-ups by only typechecking the final term once. 

\subsection{Engineering Proof Terms}
\label{sec:eng-terms}

While Coq proofs are often thought of as tactic scripts, the final product of Coq proving is \emph{proof terms} in a core type theory. Tactics merely provide a more convenient way to construct these terms (which often make heavy use of dependency). A proof that looks straightforward at the tactic level may produce proof terms that take substantially longer to check than the proof script took to run.

The central issue in term engineering is the statement of the correctness theorem. Two broad strategies are used commonly. The first, more traditional approach uses an equality proof to separate the computation from its meaning:
\begin{coq}
Theorem cancel_correct_with_eq : forall ts fs ps uvars pures l r,
  AllProvable ts fs ps uvars pures ->     (* premises (1) *)
  forall l' r', cancel l r = (l',r') ->     (* computation (2) *)
   sexprD uvars l' ===> sexprD uvars r' ->   (* denote (3) *)
   sexprD uvars l ===> sexprD uvars r.          (* denote (4) *)
\end{coq}
The benefit, and curse, of this formulation lies in the quantification over \coqe{l'} and \coqe{r'}, which together act as the result of the cancellation procedure. To apply the above theorem, our proof must directly record the values of \coqe{l'} and \coqe{r'}.
Thus, when Coq checks the proof it knows exactly what type line (3) has, regardless of the unification procedure used to justify the equation in line (2).

Representing and type checking these embedded terms in the final proof, however, can be expensive if they are large. While our representation does not necessarily look large on the surface, the dependency of terms on the type environment requires that the type environment be repeated syntactically at every constructor. Even paring down the type environment to contain only the type (eliding the equality function and its proof) does not shrink the term enough. This is a limitation of Coq's type checker, but not a theoretical one. Adapting Coq to use bi-directional type checking~\cite{pierce00bidirectional} could solve this problem. 

To circumvent the embedded term problem, we replace the quantifier with a \coqe{let} declaration scoped over the relevant premises.
\begin{coq}
Theorem cancel_correct : forall ts fs ps uvars pures l r,
  AllProvable ts fs ps uvars pures ->      (* premises (1) *)
  (let (l',r') := cancel l r in        (* computation (2) *)
   sexprD uvars l' ===> sexprD uvars r') ->   (* denote (3) *)
  sexprD uvars l ===> sexprD uvars r.            (* denote (4) *)
\end{coq}
The drawback to this approach is that na\"ive uses of this theorem do not record the result of the reduction. During proof checking, Coq will lazily evaluate this term, leaving large, partially reduced terms unevaluated at key places, requiring them to be reduced multiple times during subsequent proof checking.
While Coq provides only limited methods for specifying reduction strategies in proof terms, in practice knowing the result is enough to make proof checking reasonably efficient. In general, we can save the result by performing cut elimination with an explicit type annotation. We revisit this problem in Section~\ref{sec:eng-reduction}, discussing our solution for making this phrasing efficient.

\subsection{Engineering Reduction}
\label{sec:eng-reduction}

The workhorse of proof by reflection is the computation step, and fast evaluation of terms is essential to making reflective proofs efficient. Unfortunately, as we demonstrated in Section~\ref{sec:eng-terms}, standard formulations of reflective theorems do not work well in our setting. Our formulation requires fast \emph{delimited} evaluation, i.e. evaluation keeping certain symbols opaque. Since Coq's logic admits reduction under binders, evaluation strategies are able to handle opaque terms.  However, to maintain abstraction by not unfolding user-defined logical symbols, we need the ability to engineer the handling of particular identifiers.
Throughout our development we evaluated several reduction mechanisms, two of which are noteworthy.

\paragraph{Delimited cbv}
Coq's full-beta, call-by-value reduction mechanism is reasonably fast and supports the abstraction that we need to avoid the reduction of certain terms. As an illustration, consider the goal after an application of \coqe{cancel_correct}.
\begin{coq}
let (l',r') := cancel funcs l r in
sexprD funcs uvars l' ===> sexprD funcs uvars r'
\end{coq}
We would like to reduce the above proposition to a heap implication using the standard separation logic connectives, e.g. $*$ and \coqe{emp}. Since we know the definition of \coqe{cancel}, we can customize Coq's \coqe{cbv} tactic to leave certain constants, like $*$, opaque by specifying a whitelist of identifiers that should be reduced.

Unfortunately, specifying this list modularly is not possible. The customization available to \coqe{cbv} requires an explicit whitelist (or blacklist) of identifiers, with no facility for supporting dynamically constructed lists or using wildcards to include all values from a module. 
The whitelist required to reduce the above goal contains roughly 450 symbols, is sensitive to any refactoring (including adding additional provers), and is very difficult to debug.  
Missing symbols cause evaluation to get stuck producing enormous, partially evaluated terms that can chew through 8G of RAM in under a minute. Nevertheless, once the whitelist is complete, evaluation is fast. We post-process the resulting term \emph{folding} named definitions that may have occurred in the reflective code and in the user function environment.  That is, we substitute an identifier for its associated definition, to simplify the term.

It is this final point that makes a blacklist unattractive. Though the list would be considerably shorter, we would not know which terms to refold or how to refold them at the end of the reduction.

\paragraph{vm\_compute}

\coqe{vm_compute}~\cite{VmComputeICFP02} is an even faster Coq reduction mechanism based on compiling Gallina terms to the OCaml virtual machine, executing them there, and translating the results back to Gallina. 
The price for this speed is reduced flexibility in two ways:
\begin{enumerate}
\item Neither whitelists nor blacklists are supported.  All identifier are unfolded if definitions for them exist.
\item \coqe{vm_compute} fails if it encounters Coq unification variables.
\end{enumerate}

On the surface, both limitations appear to be show-stoppers. Without a whitelist (or blacklist), separation logic abstractions (in addition to simple functions like \coqe{plus}) will be torn apart, revealing symbolic memories and 32-bit words. Further, the goals fed to our tactics routinely contain several unification variables introduced by program-specific Ltac code.

Our solution to these problems relies on building an anonymous function that explicitly abstracts over terms that should not be reduced. This allows us to reduce only in the anonymous function and leave the dangerous subterms alone.  Consider the following simplified example, where we want to reduce the lefthand side to the righthand side:
$$
2 * 9 \Longrightarrow 9 + 9 + 0
$$
Na\"ively using \coqe{vm_compute} produces $18$. It is easy to abstract $9$, but $+$ does not occur syntactically in the term. To expose it, we use a special form of $*$ that takes $+$ as an argument:
\begin{coq}
Fixpoint mult' (plus' : nat -> nat -> nat) (n m : nat) : nat :=
  match n with 
    | 0 => 0
    | S n' => plus' m (mult' plus' n' m)
  end.
\end{coq}
Using this definition, we can engineer the following reduction: 
$$
(\mathsf{fun}\, p\, x \Rightarrow mult'\, p\, 2\, x) \overset{vm}{\Longrightarrow} (\mathsf{fun}\, p\,x \Rightarrow p\, x\, (p\, x\, 0))
$$
which differs from our target reduction by a single, cheap $\beta$-reduction when applied to $+$ and 9.

This technique works to make many terms opaque while still avoiding the problems of embedding the intermediate syntactic representation in the proof term (see Section~\ref{sec:eng-terms}); however, there are several limitations. First, we cannot use it to limit reduction in types since abstracting by a type will often produce an ill-typed term. Second, the abstractions are manifest in the resulting proof term, making it larger than it would be using delimited \coqe{cbv}. Finally, producing the abstraction using Ltac can be expensive since each abstraction must perform a linear walk over the term. 
Because we need to abstract all terms in the function and separation predicate environments, it is not uncommon to be blacklisting 30 or more symbols, each of which requires a linear pass over the term. To improve efficiency, we packaged the functionality with \coqe{vm_compute} into a tactic called \coqe{evm_compute} in a Coq plugin available online~\footnote{\url{https://github.com/braibant/evm_compute}}. Unlike \coqe{cbv}, our tactic supports dynamic blacklists by accepting Coq lists of identifiers. 


\paragraph{Comparison}
\begin{figure}
  \centering
  \scalebox{0.6}{\includegraphics{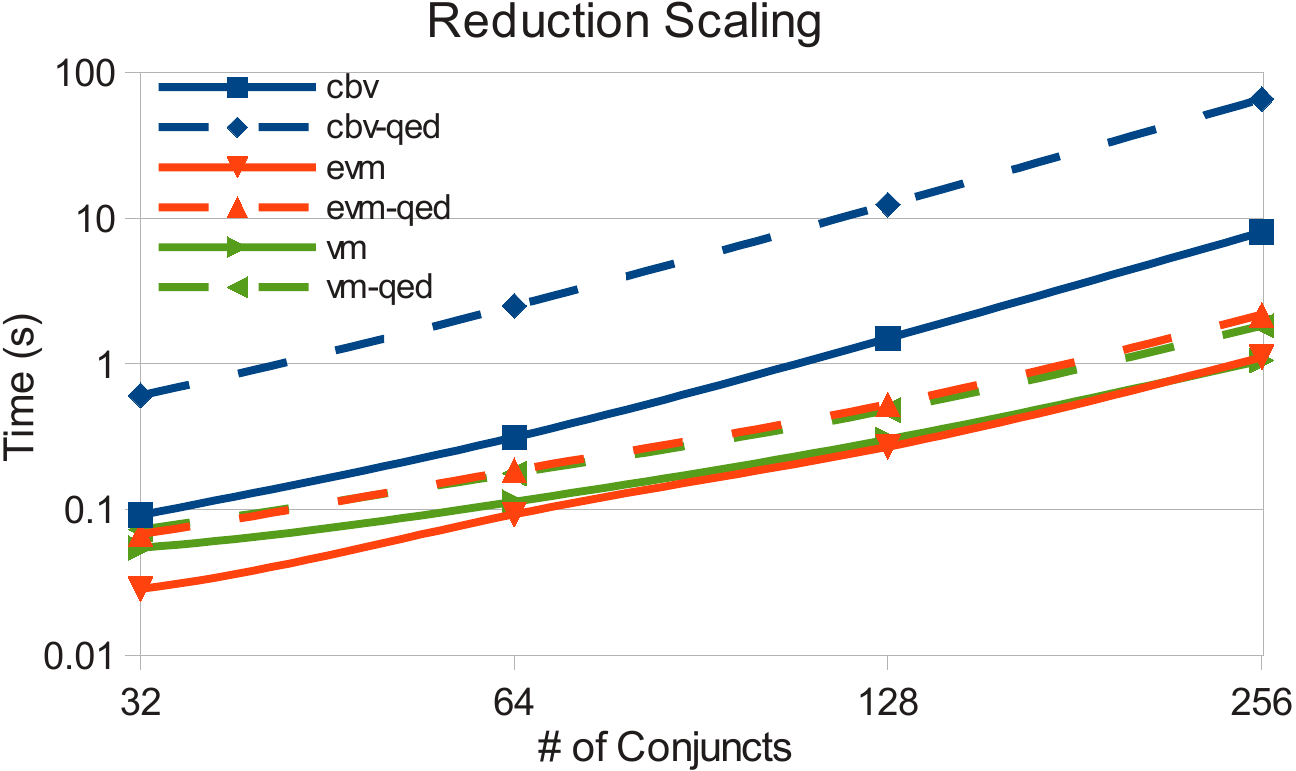}}

  \label{fig:perf-reduce}
  \caption{Performance measurements for different reduction strategies. Number of conjuncts proxies complexity.}
\end{figure}

To make the performance difference concrete, Figure~\ref{fig:perf-reduce} shows the reduction and proof checking times for delimited \coqe{cbv}, \coqe{evm_compute}, and \coqe{vm_compute}. We use the number of conjuncts as a proxy for the amount of computation since the cancellation algorithm is O($n^2$) in this case.

First, note that \coqe{cbv} is considerably slower than the virtual machine-based strategies for large problems. The further slow-down during checking \coqe{cbv} proofs is due to the customized reduction strategy not being recorded in the proof term. This causes the proof checker to fall back on lazy evaluation, which is considerably slower.

The virtual-machine based strategies are considerably faster with much better scaling properties. The overhead of blacklisting is roughly constant, becoming negligible for problems with more than 64 conjuncts. This behavior justifies the efficiency of the lightly dependent design that MirrorShard advocates. However, better facilities for customizing reduction would still be beneficial. One promising idea is to use a delimiting function such as:
\begin{coq}
Definition block (T : Type) (v : T) : T := v.
\end{coq}
While such a term would have no effect on the logical meaning of a statement, certain reduction strategies could treat occurrences of \coqe{block x} opaquely, not unfolding \coqe{x}. This would enable blocking reduction inside types and avoid the need to write functions like \coqe{mult'} in Section~\ref{sec:eng-terms} that abstract their dependencies to make them visible at the top level. Reduction strategies like \coqe{vm_compute} could then be parameterized by a set of these blocking functions. 

\section{Evaluation}
\label{sec:evaluation}

In this section we evaluate reflective proof techniques, comparing them to the standard Ltac-style verify-and-check approach to mechanized verification. We begin with a brief discussion of the automation level of our verification framework by discussing our test suite, before focusing on two grounds for comparison with Ltac-based verification methodologies: performance and debuggability.

\paragraph{Usability \& Automation}
\begin{figure}
  \begin{tabular}{l|r|r|r|r|r}
    File & Program & Invar. & Tactics & Other & Overhead \\
    \hline
    LinkedList & 42 & 26 & 27 & 31 & 2.0 \\
    Malloc & 43 & 16 & 112 & 94 & 5.2 \\
    ListSet & 50 & 31 & 23 & 46 & 2.0 \\
    TreeSet & 108 & 40 & 25 & 45 & 1.0 \\
    Queue & 53 & 22 & 80 & 93 & 3.7 \\
    Memoize & 26 & 13 & 56 & 50 & 4.6 \\
  \end{tabular}
  \caption{Case study verifications, with data on annotation burden, in lines of code}
  \label{fig:annotations}
\end{figure}
In addition to the example excerpted in Figure~\ref{treeset}, we have carried out a number of other library module verifications, to validate the usefulness and extensibility of MirrorShard. Figure~\ref{fig:annotations} shows some statistics of our six largest case studies.
In order, the columns of Figure \ref{fig:annotations} count the executable part of the module being verified, the function specifications and invariants asserted in code, the Ltac tactic proof scripts (including commands to register hints), all the remaining lines, and finally the ratio of verification lines to program lines.  
The lines that we account for under ``Other'' are almost all definitions of data structure representation predicates and statements of theorems about them.  

Our case studies exercise reasoning about a variety of user-defined abstract predicates. With the exception of a small set of obligations about words (mostly pertaining to memory access), the correctness side conditions (such as theorems about lists and sets) are verified by Ltac proof search. Our case studies are: LinkedList, consisting of the classic functions is-empty, length, reverse, and concatenate (the latter two performed in-place with mutation); Malloc, a na\"ive memory allocator, based on an unsorted free list with no coalescing, used by all the later case studies; ListSet and TreeSet, implementations of a common finite set interface specified with mathematical sets, respectively using unsorted lists and binary search trees; Queue, a standard FIFO queue specified mathematically using bags; and Memoize, a higher-order function that memoizes Bedrock code that implements a mathematical function. The last of these requires interesting interplay between our automation and custom Ltac code to handle higher-order proof obligations related to first-class code pointers.

The proof overhead is slightly lower than with the case studies used for the old fully Ltac-based Bedrock~\cite{BedrockPLDI11}. The decrease arises mostly from our modularization of hint databases. Our ability to verify the same examples demonstrates that we have achieved a similar level of automation and integration with Ltac. 

Our procedures have also been used in a larger case study~\cite{MtisSubmitted} that has built a verified cooperative threading library and then verified a Web server running on top of the library.  The thread library includes about 400 lines of implementation code and 3000 additional lines for its verification, while the Web server has 200 lines of implementation and 500 more for the proof, which establishes that representation invariants are maintained for key data structures.

\paragraph{Performance}
Beyond expressive power, a crucial benchmark for verification tools is performance. Long-running tactics (or tools) cut the programmer out of the loop, making iterative development difficult. 

\begin{figure}
  \center
  \scalebox{0.54}{\includegraphics{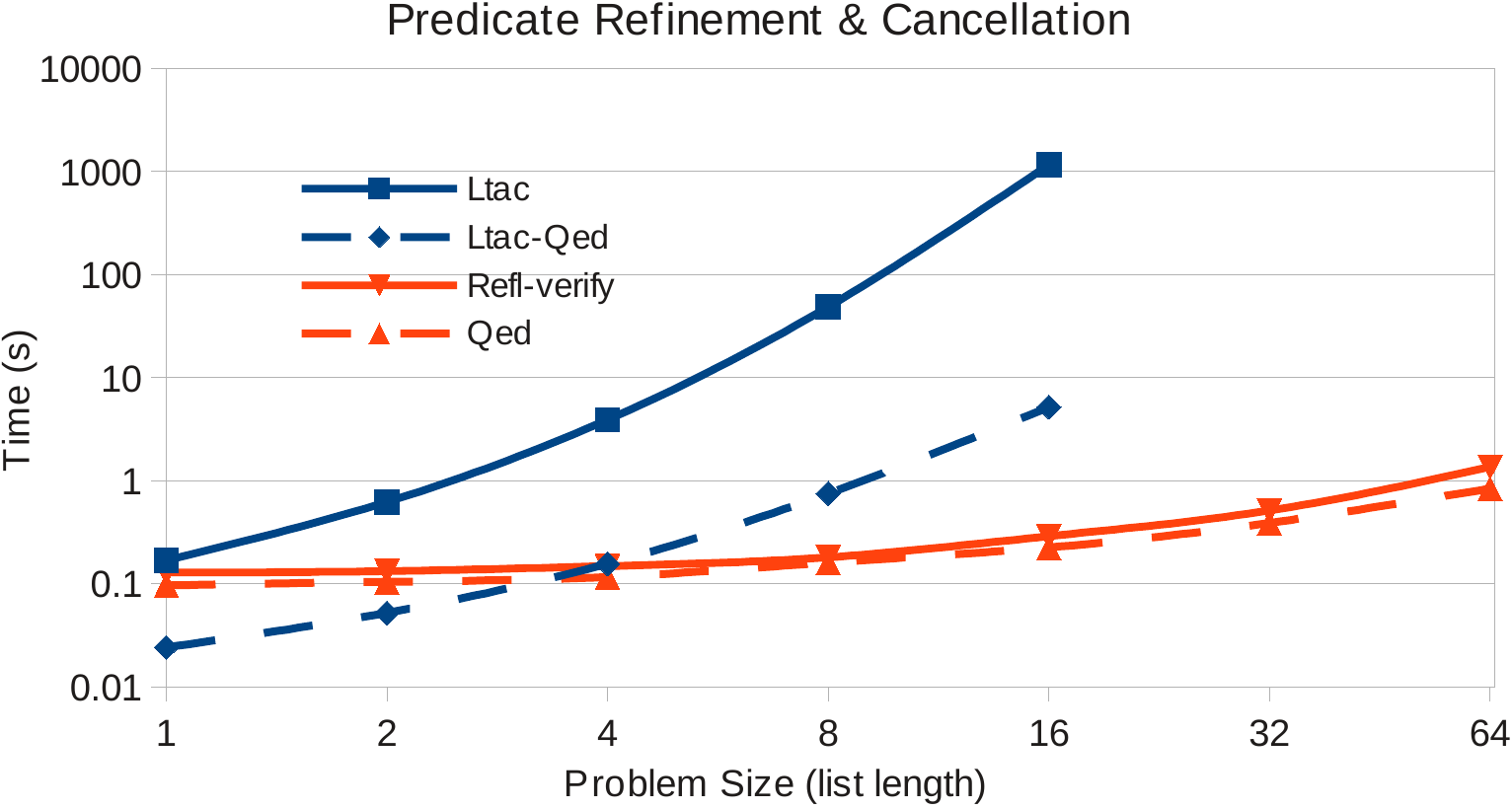}}
  \label{fig:performance}
  \caption{Performance comparison to non-reflective procedures}
\end{figure}

Figure~\ref{fig:performance} uses a microbenchmark to compare the performance of our reflective procedures based on MirrorShard to those Chlipala developed for his initial version of Bedrock~\cite{BedrockPLDI11}.  The background for this task is an abstract predicate $\mathsf{sll}$ for singly linked lists, along with two \emph{theorems} that we use as refinement hints:

{\small
$$\lceil p = 0 \rceil \Longrightarrow \mathsf{sll}([], p)$$
$$\lceil p \neq 0 \rceil * \exists p'. \; p \mapsto x, p' * \mathsf{sll}(\ell, p')) \Longrightarrow \mathsf{sll}(x :: \ell, p)$$
}

Out of these theorems, we can derive variants for concrete list lengths.
For readability, we leave out side conditions on nullness or non-nullness of pointer variables, which appear in our actual benchmark theorem statements.
\begin{eqnarray*}
  \mathsf{emp} &\Longrightarrow& \mathsf{sll}([], p_0) \\
  p_0 \mapsto x_0, p_1 &\Longrightarrow& \mathsf{sll}(x_0 :: [], p_0) \\
  p_0 \mapsto x_0, p_1 * p_1 \mapsto x_1, p_2 &\Longrightarrow& \mathsf{sll}(x_0 :: x_1 :: [], p_0) \\
\end{eqnarray*}
\noindent ...and so on, generalizing to an arbitrary number of list cells.  If the length of the list is $n$, solving this problem requires $n+1$ refinements, with $n$ refinements via the theorem for non-empty lists and the final refinement using the empty-list theorem.  In the process, we introduce $n$ unification variables and $n$ pure facts (that none of the intermediate pointers are equal to 0).

Proving this family of theorems using the Ltac automation from the old Bedrock system~\cite{BedrockPLDI11} is painfully slow both to find a proof (Ltac) and check it (Ltac-Qed).  Our experiments time out for a list of length 32, while our new reflective automation (Refl) finishes in under a second.  We also see that the reflective tactic spends only slightly longer on proof search than checking, while with the old Ltac approach we see proof search running for at least 10 times longer.  It is now faster to \emph{find} proofs than it had been to \emph{check} them.

This straightforward result is the ``good news'' arising from our experiments.  We achieve asymptotically better performance scaling than the Ltac-based alternative, and the constants are low enough that the performance gap becomes clear even for relatively small microbenchmarks.

The ``bad news'' arising from our experiments is that we see no clear change in overall performance for our full case studies like in Figure \ref{fig:annotations}.  Our experimental set-up is quite conservative, since our new Bedrock system involves a number of complexities not found in the original.  For instance, we added support for higher-order quantification in assertions; we switched the machine word representation from natural numbers to size-32 bitvectors; and we introduced the possibility for programs to \emph{crash} by accessing invalid memory addresses, creating many new crash-safety proof obligations for each program.  Seen from this perspective, one might consider it a very promising sign that we hold overall verification performance at approximately the same level.  It is probably also true that programs making good use of data structure encapsulation will tend to feature relatively small assertions that do not provide much opportunity to show off the asymptotic scaling of proof procedures.

\begin{figure}
  \center
  \scalebox{0.4}{\includegraphics{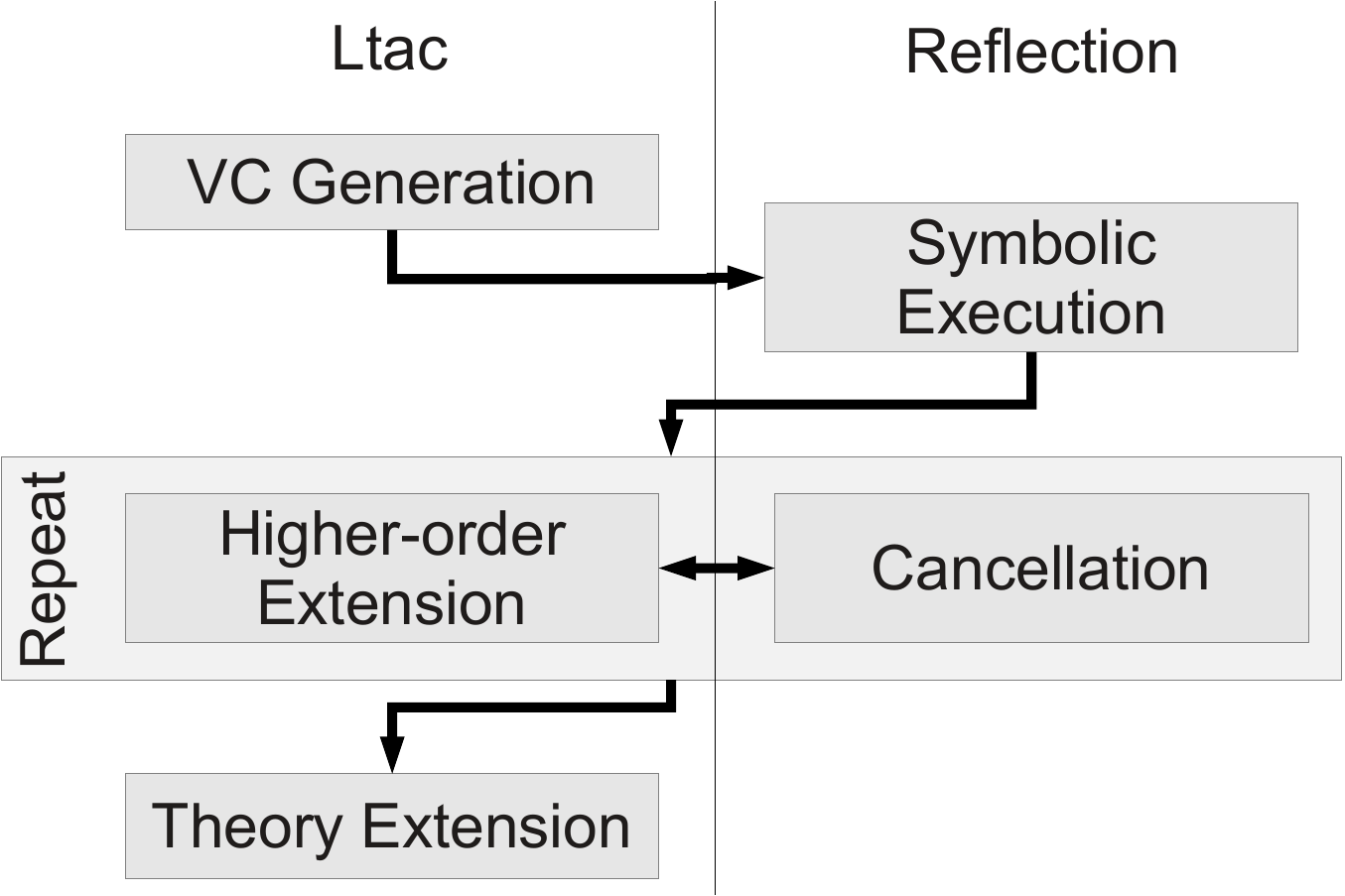}}

  \vspace{0.2cm}

  \scalebox{0.63}{\includegraphics{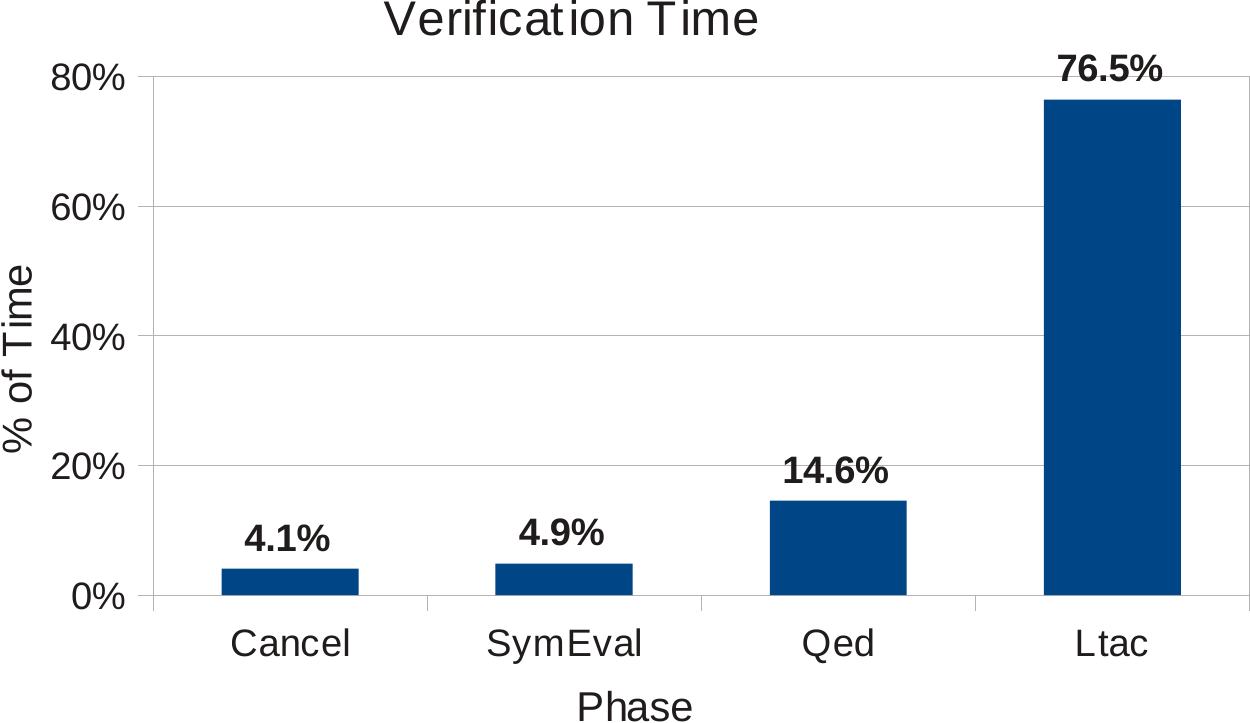}}
  
  \caption{Verification process and the breakdown of verification time}
  \label{fig:process}
\end{figure}
Figure~\ref{fig:process} shows how our reflective proofs fit into the overall verification. The flexibility afforded to us by this method is, in some sense, its downfall. Two-thirds of verification time is spent in Ltac, and pushing more reasoning into Gallina procedures is likely to reduce verification time drastically. We expect that our general techniques to support quantifiers and integrate pure provers should streamline further development of similar procedures.  As we experiment with more programs to verify, we expect both to improve the performance of our pure provers, by introducing more efficient Gallina data structures; and to add new procedures to discharge obligations in new mathematical theories.

\paragraph{Debuggability}
A crucial benefit of reflective proofs over their Ltac counterparts beyond performance is the ability to reason about the correctness of the proof-generating procedure. Ltac programs have complicated backtracking semantics that can make them difficult to write and even more difficult to debug. For example, the backtracking severely complicated debugging our Ltac-based reification code, since a typo in a single case would cause an exponential backtrack through the algorithm. In addition, debugging tools are difficult, and tactics that compute terms must be hand-coded in continuation-passing style to get reasonable debugging support via \coqe{idtac}, Ltac's equivalent of \coqe{printf} debugging. 

On the other hand, even with a minimally dependent term representation, coding in Gallina enables us to use Coq's type checker to get shallow ``sanity'' properties. Our soundness theorems allow us to prove the deeper properties that we are relying on. During development we found ourselves frequently fixing bugs related to de Bruijn indices and binders, up \emph{until} the point when the proofs about the components were completed. The proving process contributed considerably to the development process, pointing out bugs that our initial test cases did not cover.

\section{Related Work}
\label{sec:related-work}


Our work is part of a recent trend to improve the automation available in proof assistants, which have traditionally supported only very manual proof styles. Researchers have proposed several alternative approaches.

Proof by reflection~\cite{ReflectionTACS97} is a well-established technique in the communities of Coq and other closely related proof assistants.  Gr\'egoire and Mahboubi built a reflective tactic to simplify terms using the operators of any \emph{ring} algebraic structure~\cite{RingTPHOLs05}, and Braibant and Pous built a reflective implementation of rewriting modulo associativity and commutativity of user-specified operators~\cite{AcCPP11}.  These past projects consider self-contained, well-defined problems in the style of classical decision problems.  In contrast, our work considers \emph{open-ended, extensible} procedures more along the lines of those commonly used for automated program verification.  Such an expansion of scope raises the new issues that we have described, like supporting quantifiers, an interface with a proof assistant's unification engine, and modular combination of verified decision procedures over different theories.  The last of these has been considered by Lescuyer~\cite{lescuyer11these}, who developed an SMT implementation in Coq. The theory composition that he achieves is more integrated than the simple composite provers that we implemented, but the approach does not share the computational composition that enables us to achieve lightweight extension. The Ssreflect Coq library~\cite{SsreflectJFR10} employs a \emph{small-scale reflection} style where many predicates are coded as functional programs returning Booleans, sidestepping concerns of decidability.  The approach of Gonthier et al.~\cite{AdHocICFP11} uses Coq's canonical structures mechanism as a clever means of building proof-generating procedures, retaining most of the usual relative advantages and disadvantages of proof generation versus verification of proof procedures like ours.

Other recent work has proposed Mtac~\cite{mtac}, a new style of proof automation in Coq. Like reflective proofs, Mtac proof procedures are implemented in Gallina; however, in order to to provide the types of operations necessary for making tactic development simple, these ``tactics'' have a monadic type. The monad supports non-termination, failure, and syntactic matching of patterns against terms. The last of these features makes it impossible to reason about Mtac procedures inside of Coq, since syntactic matching breaks the substitution property of equality, i.e. $x = y \rightarrow f x = f y$. Execution of these programs is done at type-elaboration time through a special \texttt{run} expression that exists outside of Gallina.

Several projects~\cite{CvcPDPAR05,Smt3CPP11} have studied translation of SMT-solver proof traces into forms acceptable to proof assistants, and some of these projects~\cite{Smt1CPP11,Smt2CPP11} are based on reflective Coq tactics.  In the latter case, one verifies a \emph{proof checker} rather than the prover itself.  Compared to our approach, there are non-obvious performance trade-offs. Verifying the prover removes the need for potentially expensive proof generation and checking, but the proof-generating approach is compatible with using efficient low-level languages and optimizing compilers to implement the provers.  Verifying the prover helps avoid completeness bugs, where a tool may sometimes generate invalid proof traces; but proof checkers are generally easier to verify than provers.  Lescuyer and Conchon~\cite{SatFroCos09} built a reflective SAT solver tactic for Coq, and Nanevski et al.~\cite{HttPOPL10} and Oe et al.~\cite{VersatVMCAI12} have verified efficient low-level code for a part of an SMT solver and a full SAT solver, respectively.  None of this past work supports modular extension with new \emph{provers} rather than just \emph{proof checkers}, and none supports a rich formula language including quantifiers and user-specified predicates with associated axioms that should be applied automatically.

A few past projects have proved the correctness of non-extensible separation logic proof procedures.  Marti and Affeldt~\cite{AffeldtCS08} verified a simplification of Smallfoot~\cite{SmallfootFMCO05} using Coq.  Stewart et al. have done Coq verification of a Smallfoot-style verification tool VeriSmall~\cite{VeriSmallCPP11} that relies on a novel verified heap theorem prover VeriStar~\cite{VeriStarICFP12}.  The prior work of this kind has considered none of functional correctness verification (as opposed to just memory safety), extension with abstract predicates, or higher-order programs or specifications.

Many standalone tools do efficient, automated analysis of large low-level code bases for memory safety, using separation logic, outside of the context of proof assistants.  Examples include Smallfoot~\cite{SmallfootFMCO05}, SpaceInvader~\cite{SpaceInvaderPOPL09}, and SLAyer~\cite{SLAyerCAV11}.  Xisa~\cite{XisaPOPL08} bears a special relationship to our new work, as it is extensible with new predicate definitions in separation logic.  Several other proof assistant libraries provide support for separation logic proofs, including the tactic libraries of Appel~\cite{AppelTactics} and McCreight~\cite{PtslTPHOLs09}, Holfoot~\cite{HolfootTPHOLs09}, Ynot~\cite{YnotICFP09}, and Charge!~\cite{ChargeITP12}.  Some of the libraries in this latter category provide proof automation comparable to that of the standalone tools.  Our work described in this paper is the first to verify such automation formally, rather than merely constructing it to output program-specific proofs.  One disadvantage of all such approaches is greater performance overhead compared to standalone tools, though traditional proof techniques can be applied directly in places where sophisticated, custom reasoning is necessary. 

\section{Conclusion}
\label{sec:conclusions}

We have built a core reflective proof framework that supports user extensions via reflected lemmas and custom proof procedures that are packaged into reusable strategies. Our framework allows bi-directional communication with Coq's unification variables, supporting both the instantiation of existing unification variables and the construction of new unification variables.

To justify the framework's applicability, we have instantiated it to reason about a combination of higher-order and separation logic including support for user-defined abstract predicates. In addition to the benefit of user extension for handling abstract predicates, our reflective tactics scale much better performance-wise than Bedrock's original tactic-based verification procedure~\cite{BedrockPLDI11}, while producing at least comparable performance for the realistic case studies that we have experimented with.

We succeeded in isolating two large chunks of separation logic-based verification engines. Looking forward, it seems that the ripest areas for performance improvements are the non-reflective portions. Building a larger library of (reusable) base theory provers in our framework would reduce the number of goals passed back to Ltac and enable us to apply more refinements reflectively. It would also be interesting see how the framework can be extended to capture fragments of higher-order logic that are common during verification. This would enable us, in simple cases, to avoid some of the ping-ponging which we believe is the source of much of our overhead. While full reflective verification would be ideal, the ability of our framework to integrate nicely with more manual proofs enables us to choose to invest time in automation only when there will be a comparable payoff, i.e. where similar obligations crop up repeatedly. Highly specialized reasoning can still be proved manually or semi-automatically.
\bibliographystyle{abbrvnat}

\bibliography{bedrock.bib}

\end{document}